\documentclass[fleqn,usenatbib]{mnras}

\usepackage{newtxtext}

\usepackage{amssymb}

\usepackage{newtxmath}

\usepackage[T1]{fontenc}
\usepackage[table,xcdraw,dvipsnames]{xcolor}
 \usepackage{multicol}
\DeclareRobustCommand{\VAN}[3]{#2}
\let\VANthebibliography\thebibliography
\def\thebibliography{\DeclareRobustCommand{\VAN}[3]{##3}\VANthebibliography}
\usepackage[colorinlistoftodos]{todonotes}

\newcommand{\red}[1]{\comm} 

\usepackage{graphicx}	
\usepackage{amsmath}	
\usepackage{amssymb}	

\title[HTRU-S low lat with the FFA pipeline]{The High Time Resolution Universe Pulsar Survey – XVIII. The reprocessing of the HTRU-S Low Lat survey around the Galactic centre using a Fast Folding Algorithm pipeline for accelerated pulsars }

\author[J.  Wongphechauxsorn et al.]{
J. Wongphechauxsorn ,$^{1}$\thanks{E-mail: jompoj@mpifr-bonn.mpg.de}
D. J. Champion,$^{1}$
M. Bailes,$^{2,3}$
V. Balakrishnan,$^{1}$
E. D. Barr,$^{1}$
\newauthor
M. C. i Bernadich,$^{1}$
N. D. R. Bhat,$^{4}$
M. Burgay,$^{5}$
A. D. Cameron,$^{2,3}$
W. Chen,$^{1}$
C. M. L. Flynn,$^{2,3}$
\newauthor
A. Jameson,$^{2,3}$
S. Johnston,$^{6}$
M. J. Keith,$^{7}$
M. Kramer,$^{1,7}$
C. Ng,$^{8}$
A. Possenti,$^{5,9}$
R. Sengar,$^{10, 2,3}$
\newauthor
R. M. Shannon,$^{2,3}$
B. Stappers,$^{7}$
W. van Straten$^{10}$
\\
$^{1}$Max-Planck-Institut für Radioastronomie, Auf dem Hügel 69, D-53121 Bonn, Germany\\
$^{2}$ARC Center of Excellence for Gravitational Wave Discovery (OzGrav), Swinburne University of Technology, Mail H11, PO Box 218, VIC 3122.\\
$^{3}$Centre for Astrophysics and Supercomputer, Swinburne University of Technology, P.O. Box 218, Hawthorn, Victoria 3122, Australia\\
$^{4}$International Centre for Radio Astronomy Research, Curtin University, Bentley, WA 6102, Australia.\\
$^{5}$ INAF - Osservatorio Astronomico di Cagliari, Via della Scienza 5, I-09047 Selargius (CA), Italy.\\
$^{6}$CSIRO Astronomy $\&$ Space Science, Australia Telescope National Facility, P.O. Box 76, Epping, NSW 1710, Australia.\\
$^{7}$Jodrell Bank Center for Astrophysics, University of Manchester, Alan Turing Building, Oxford Road, Manchester M13 9PL, United Kingdom.\\
$^{8}$Dunlap Institute for Astronomy \& Astrophysics, University of Toronto, 50 St.~George Street, Toronto, ON M5S 3H4, Canada.\\
$^{9}$Universit´a di Cagliari, Dept of Physics, S.P. Monserrato-Sestu Km 0,700 - 09042 Monserrato, Italy.\\
$^{10}$Center for Gravitation, Cosmology, and Astrophysics, Department of Physics, University of Wisconsin-Milwaukee, P.O. Box 413, Milwaukee, WI 53201, USA. \\
$^{11}$Institute for Radio Astronomy $\&$ Space Research, Auckland University of Technology, Private Bag 92006, Auckland 1142, New Zealand.
}

\date{Accepted XXX. Received YYY; in original form ZZZ}

\pubyear{2021}

\begin{document}
\label{firstpage}
\pagerange{\pageref{firstpage}--\pageref{lastpage}}
\maketitle

\begin{abstract}

The HTRU-S Low Latitude survey data within 1$^{\circ}$of the Galactic Centre (GC) were searched for pulsars using the Fast Folding Algorithm (FFA). Unlike traditional Fast Fourier Transform (FFT) pipelines, the FFA optimally folds the data for all possible periods over a given range, which is particularly advantageous for pulsars with low-duty cycle. For the first time, a search over acceleration was included in the FFA to improve its sensitivity to binary pulsars. The steps in dispersion measure (DM) and acceleration were optimised, resulting in a reduction of the number of trials by 86 per cent. This was achieved over a search period range from 0.6-s to 432-s, i.e. 10 per cent of the observation time (4320s), with a maximum DM of 4000 pc cm$^{-3}$ and an acceleration range of $\pm 128$m s$^{-2}$. The search resulted in the re-detections of four known pulsars, including a pulsar which was missed in previous FFT processing of this survey. This result indicates that the FFA pipeline is more sensitive than the FFT pipeline used in the previous processing of the survey within our parameter range. Additionally, we discovered a 1.89-s pulsar, PSR J1746-2829, with a large DM, located~0.5 from the GC. Follow-up observations revealed that this pulsar has a relatively flat spectrum($\alpha=-0.9\pm0.1$) and has a period derivative of $\sim1.3\times10^{-12}$ s s$^{-1}$, implying a surface magnetic field of $\sim5.2\times10^{13}$ G and a characteristic age of $\sim23000$ yr. While the period, spectral index, and surface magnetic field strength are similar to many radio magnetars, other characteristics such as high linear polarization are absent.

\end{abstract}




\section{Introduction}

The Galactic Centre (GC) is regarded as one of the most interesting regions in our Galaxy due to the dense environment of matter and relatively large magnetic field around the supermassive black hole, Sgr A* \citep{1996Natur.383..415E}. Discovering radio pulsars, a subclass of neutron stars whose rotation is visible in pulses of radio emission detectable at Earth, located in the GC can lead to many applications, including studying stellar evolution and the magneto-ionised environment around the GC \citep[e.g.][]{2018ApJ...852L..12D,2023MNRAS.524.2966A}. Moreover, the discovery of a typical pulsar orbiting Sgr A* with an orbital period of approximately one year would be sufficient to test two predictions of General Relativity Theory \citep{wex1999frame,2004NewAR..48..993K,liu2012prospects,2017NatAs...1..812L}: the No-Hair-Theorem \citep[e.g.][]{  1967PhRv..164.1776I,1968CMaPh...8..245I} and the Cosmic Censorship Conjecture \citep[e.g.][]{Penrose:1980ge}. The No-Hair-Theorem predicts that all stationary black holes can be described by their mass, spin, and electric charge alone, while the Cosmic Censorship Conjecture predicts that all black holes must be surrounded by event horizons.

The environment around the GC favours the formation of massive stars \citep[see e.g.][for a review]{2003ANS...324..255F}. It is predicted that there will be $10^7$-$10^8$ neutron stars, with 10$^{2}$ to 10$^{6}$ of them being pulsars at the central 100 pc \citep{cordes1997finding}. \citet{wharton2012multiwavelength} estimated that the inner 150 pc of the Galaxy could harbour as many as 1000 active radio pulsars that beam towards Earth, further motivating ongoing searches for pulsars around the GC. However, only six pulsars have thus far  been detected within half a degree of angular separation from the GC \citep{johnston2006discovery,deneva2009discovery,2013Natur.501..391E}. 

One reason for the disparity between the predicted and observed number of pulsars is that the GC environment diminishes pulsar detectability. The dense environment in the GC affects the pulsar's detectability in two main ways. Firstly, the arrival times of the radio pulses are delayed towards lower observational frequency as $\Delta t \propto f^{-2}$. This effect, known as `dispersion', originates from the ionised interstellar medium (ISM), and can be entirely mitigated by performing de-dispersion (see Section \ref{DM}. Secondly, multi-path propagation caused by scattering in the ISM broadens the pulse width with a timescale that scales as $\tau_{\rm ts} \propto f^{-4}$ for a thin screen scenario \citep[see][for a review]{1977ARA&A..15..479R}. Critically, this effect can only be reduced by {observing} {at} higher frequencies. If the broadened pulse width is larger than the pulse period, it is impossible to detect periodicity from the pulsar as the pulses are completely smeared out. At the GC, the scatter broadening time of radio pulses at 1.4 GHz is expected to be as large as 2300 s \citep[e.g.][]{2002astro.ph..7156C}, longer than known periods of pulsars \citep[e.g. \textsc{PSRcat},][]{2005AJ....129.1993M}\footnote{https://www.atnf.csiro.au/research/pulsar/psrcat/\label{psrcat}}. This is one possible explanation for the lack of pulsar discoveries around the GC from the numerous pulsar surveys to date \citep[e.g.][]{1996ASPC..105...13K,2004IAUS..218..133K,2022ApJ...933..121S,iram,wharton_2017,2021ApJ...914...30L,eatough2021multi}. By comparison, real data shows that the scattering broadening of the GC magnetar PSR ~J1745$-$2900, the closest pulsar to Sgr A* with a projected distance of 0.1 pc from the GC \citep{2013Natur.501..391E,2013ApJ...770L..24K}, is only on the order of seconds \citep{spitler2013pulse}. This is in stark contrast to the previous prediction. However, even this lower scattering time may still be responsible for the scarcity of pulsars detected, given that this scattering time at the typical pulsar search frequency, 1.4 GHz, is longer the period of approximately 30 per cent of the known pulsars  \citep[e.g. \textsc{PSRcat},][]{2005AJ....129.1993M}\footref{psrcat}.

Conducted over the last 25 years, most pulsar surveys have used an observing frequency around 1.4 GHz. This balances the reduction in sensitivity due to scatter broadening at low frequencies with the steep pulsar spectrum \citep[see e.g. ][]{spectral0,spectral}. The anomalously low number of pulsars in this region suggests more substantial chromatic effects \citep[see][for example]{2017MNRAS.471..730R}, prompting surveys at higher frequencies. High frequency pulsar surveys at the GC range from 4.0-8.0 GHz using the 100-m Effelsberg, 100-m Green bank, and 64-m Parkes (Murriyang) telescopes  \citep{1996ASPC..105...13K,2004IAUS..218..133K,2010ApJ...715..939M,2022ApJ...933..121S} to 84 and 156 GHz \citep{iram} using the 30-m IRAM telescope. To reduce the effect of red-noise caused by fluctuations in the telescope's receiver electronics during long integration observations, GC surveys have also been conducted using interferometers e.g., the Atacama Large Millimeter/submillimeter Array (ALMA) and Very Large Array (VLA) \citep{wharton_2017,2021ApJ...914...30L}. In order to increase the sensitivity of the surveys as the pulsars become weaker at high frequencies, \cite{eatough2021multi} made repeated high frequency observations (4.85, 8.35, 14.6, and 18.95 GHz) on a time scale of years. The long duration of the observations meant that the apparent period changes of a potential pulsar due to orbital motion needed to be accounted for not only by including the searches for acceleration (period derivative) but also jerk (the second derivative) (see Section \ref{acc}). Despite these numerous higher-frequency pulsar surveys with various observation techniques, no new pulsars have been found.

Rather than competing with the spectral index of the pulsars in the GC, another way to reduce the deleterious effect of scattering is to look for longer period pulsars. Previous searches were biased toward the fast-spinning pulsars over the slow-spinning ones through their use of the Fast Fourier Transform (FFT) algorithm to detect the repeating signal. \citet{cameron2017investigation,parent2018implementation,morello2020optimal} suggested that the reduction in the FFT sensitivity to long-period pulsars with harmonic components in the Fourier domain (e.g. a narrow pulse profile) is significantly more severe than anticipated, primarily due to factors such as the impact of red instrumental noise at the low-frequency end of the Fourier power spectrum and incoherence harmonic summing. Another reason to search for long-period pulsars in the GC is also driven by the hypothesis is \citet{2014ApJ...783L...7D} that the GC may exhibit a higher efficiency of magnetar  formation (see Section \ref{sec:mag}). In this work, we searched for long-period pulsars that would not be as strongly affected by a significant scattering of the GC using the Fast Folding Algorithm \citep[FFA,][]{staelin1969fast} which is more sensitive to long-period pulsars.

The organisation of this paper is as follows. Section \ref{Methods} describes the methods used in this work, including data selection and the algorithm. The optimisations, the pipeline, and the search parameters are further elaborated. Section \ref{Results} reports the re-detection of the known pulsars and a new pulsar discovered in this survey. The discussion of the possible radio image counterpart of the newly found pulsar and the type of newly discovered pulsar is shown in Section \ref{discussion}. Finally, we present our conclusion in Section \ref{Conclutions}.

\section{Methods}\label{Methods}
\subsection{Data selection} \label{Data}
The 64-m Parkes radio telescope (Murriyang), located in New South Wales, Australia, was used for the Southern High Time Resolution Universe Survey (HTRU-S). Observations started in 2008 and ended in 2014 and made use of the Parkes Multibeam receiver which consists of 13 feeds, producing 13 telescope beams \citep{staveley1996parkes}. The HTRU-S survey was divided into three regions based on Galactic latitude. We used only the data set that formed the low Galactic latitude survey (LowLat) in this work because it has the longest observation time (4320 s for each pointing) with a sampling time ($t_{\rm{samp},0}$) of 64 $\mu$s, containing $2^{26}$ samples per observation and covered the GC \citep{ng2015high}. The central observing frequency $f_{\rm{c}}$ is at 1352 MHz with a bandwidth ($\Delta f$) of 340 MHz, separated into effective 870  channels \citep{keith2010high}. The long observation time of LowLat also favours the discovery of slow pulsars as it records more pulses because the signal-to-noise ratio (S/N) is proportional to the observation time ($t_{\rm obs}$) as $\rm{S/N}\propto\sqrt{T_{\rm obs}}$. Additional information regarding other HTRU surveys can be found in \citet{keith2010high,ng2015high} and \citet{cameron2020high}.

Because this work is focused on the GC, observation beams within one degree of the GC were selected. The data were processed at the Max-Planck Computing and Data Facility in Garching, Germany, using the MPIfR's Hercules cluster. 

\subsection{The search algorithm}
\label{algorithm}

As mentioned in the previous section, the vast majority of pulsars have thus far been discovered by performing an FFT on a time-series and searching for signals in the Fourier spectrum; this is due to its computational efficiency. {However, narrow duty-cycle signals become more difficult to detect as more power is distributed to the higher harmonics in the Fourier domain. A fraction of the Fourier power can be recovered by performing an ``Incoherent harmonic  summing'' technique \citep{taylorTwoNewPulsating1969}. This method adds the power of each harmonic to the fundamental frequency, recovering the missing Fourier power. However, it is impossible to recover all of Fourier power if the pulse is sufficiently narrow. Furthermore,  as a result, the FFT has some limitations when applied to the detection of long-period (P>1 s) and narrow pulse pulsars (pulse width smaller than 20 per cent of the period for 32 harmonic sums) \citep[see ][]{morello2020optimal}.}

The FFA is an alternative method used to search for periodicity by brute-force folding every possible trial period. These folded pulse profiles must then be evaluated statistically, e.g. using matched-filtering techniques. Although the FFA was first implemented in 1969 \citep{staelin1969fast}, the algorithm has only been applied to a few pulsar surveys \citep[e.g.,][]{2004MNRAS.355..147F,kondratiev2009new} as it is computationally expensive. Recently, \citet{morello2020optimal} presented a new FFA implementation (hereafter the  \texttt{Riptide} FFA)\footnote{https://github.com/v-morello/riptide}. This implementation of the FFA has been demonstrated to be faster than the previous implementations and can be used in conjunction with a blind pulsar survey. It has already resulted in several discoveries, including PSR~ J0043$-$73 \citep{titus2019targeted}, a new pulsar in the Small Magellanic Cloud, and PSR~J2251$-$3711, a pulsar from the SUPERB survey \citep{keane2018survey} with a spin period of 12.1s \citep{morello2020survey}. Recently, \citep{2022ApJ...934..138S,2023ApJ...944...54S} discovered six new pulsars with the \texttt{Riptide} FFA, one of which has a period of 140-ms, using data from the GMRT High Resolution Southern Sky \citep{2016ApJ...817..130B} pulsar survey.

In this work, we searched for periodicity in the selected data using the \texttt{Riptide} FFA implementation. To avoid confusion between a general FFA implementation, the \texttt{Riptide} FFA, and the acceleration search pipeline implemented in this work, we will refer to them as FFA, RFFA, and AFFA, respectively, for the remainder of this paper. We also present a new method to optimise the dispersion and acceleration trials when searching for pulsars in binary systems. Those optimisations could be done due to the fact that the search step size is proportional to the minimum search period, as the period searches in the FFA are independent (unlike the FFT) so that they can be individually tailored. 

\subsection{Optimisation of dispersion measure and acceleration trials}
\label{optimization}
To improve the sensitivity of a survey to highly dispersed pulsars, the effect of dispersion must first be corrected for. While the FFT searches for all possible pulse frequencies simultaneously, the FFA only folds a small period range at a time. Since the acceleration and dispersion step sizes are directly proportional to the profile bin width, represented as \( t_{\text{samp}} \) in this study, and in turn, \( t_{\text{samp}} \) is proportional to the minimum search period, it is thus possible to optimise the search step size for each period range.

\subsubsection{Searched period}
\label{period}
First, we define the number of bins we want in each profile ($N_{\rm bins}$). Then we can determine the minimum search period ($P_{\rm{min}}$) for a time-series with sampling time $t_{\rm{samp}}$: 
\begin{equation}
\label{tsamp}
    P_{\rm{min}}<=N_{\rm{bins}}\times  t_{\rm{samp}}.
\end{equation}
As we search for longer periods with the same $N_{\rm bins}$ the time-series can be downsampled. Keeping the same $N_{\rm bins}$ also reduces the required computational resources. In this work, the RFFA was used to downsample the time-series by the factor of 2 whenever the search period doubles \citep[see][for more details]{morello2020optimal}.

\subsubsection{Searched dispersion measure }
\label{DM}

The dispersive time delay between two observing frequencies is proportional to the dispersion measure (DM), and is given by:
{
\begin{equation}
\label{DM_smear}
\Delta t = 4.15 \times 10^{6} {\rm ms} \times (\frac{f_1^{-2}}{\rm MHz}-\frac{f_2^{-2}}{\rm MHz}) \times \rm{\frac{DM} {pc \ cm^{-3}}}.
\end{equation}}Dispersion can be mitigated by spitting the bandwidth into channels and progressively shifting each channel in time, before summing them to create a time-series. This method is called the de-dispersion. As the DM is initially unknown, a range of DM must be searched. The $i^{\rm{th}}$ dispersion step at central frequency ($f_{\rm c}$) at bandwidth ($\Delta f$) and sampling time ($t_{\rm{samp}}$) in ms is calculated using 
{
\begin{equation}
\label{DM_step}
 {\rm DM}_i = 1.205 \times 10^{-7} \times (i-1) \times t_{\rm{samp}} \left (\frac{f_c^3}{\rm MHz} \frac{{\Delta f}^{-1}}{\rm MHz}\right ) \, \rm{pc} \, \rm{cm}^{-3},
\end{equation}}For more information, see \citet{2012hpa..book.....L} and references therein. As the FFA continuously downsamples the data to maintain the same number of profile bins, the size of the DM steps increases, reducing the total number of trials.

\subsubsection{Acceleration search}
\label{acc}
Any orbital motion will cause an apparent period change in the pulsar during the observation, depending on the relative size of the orbital period and observation length. This will smear and reduce S/N of the profile and may make the pulsar undetectable. Ideally, a search over all five Keplerian parameters would be performed \citep[e.g.][]{2022MNRAS.511.1265B}, however for a large survey this is too computationally expensive. A simpler approach is to assume that the pulsar is moving with constant acceleration along the line of sight, $a$, at a reference time $t$. \cite{1991ApJ...368..504J} demonstrated that the modulated pulse arrival time ($\Delta t$) is 
\begin{equation}
\Delta t = \frac{at^2}{2 c}
\end{equation}
where $c$ is speed of the light. \cite{ransom2003new} and \citet{ng2015high} demonstrated that this approximation is true only if the observation time is less than $10$ per cent of the orbital period due to the fact that the observation time is short enough to consider the acceleration to be constant.

\begin{figure}
 \includegraphics[width=0.95\columnwidth]{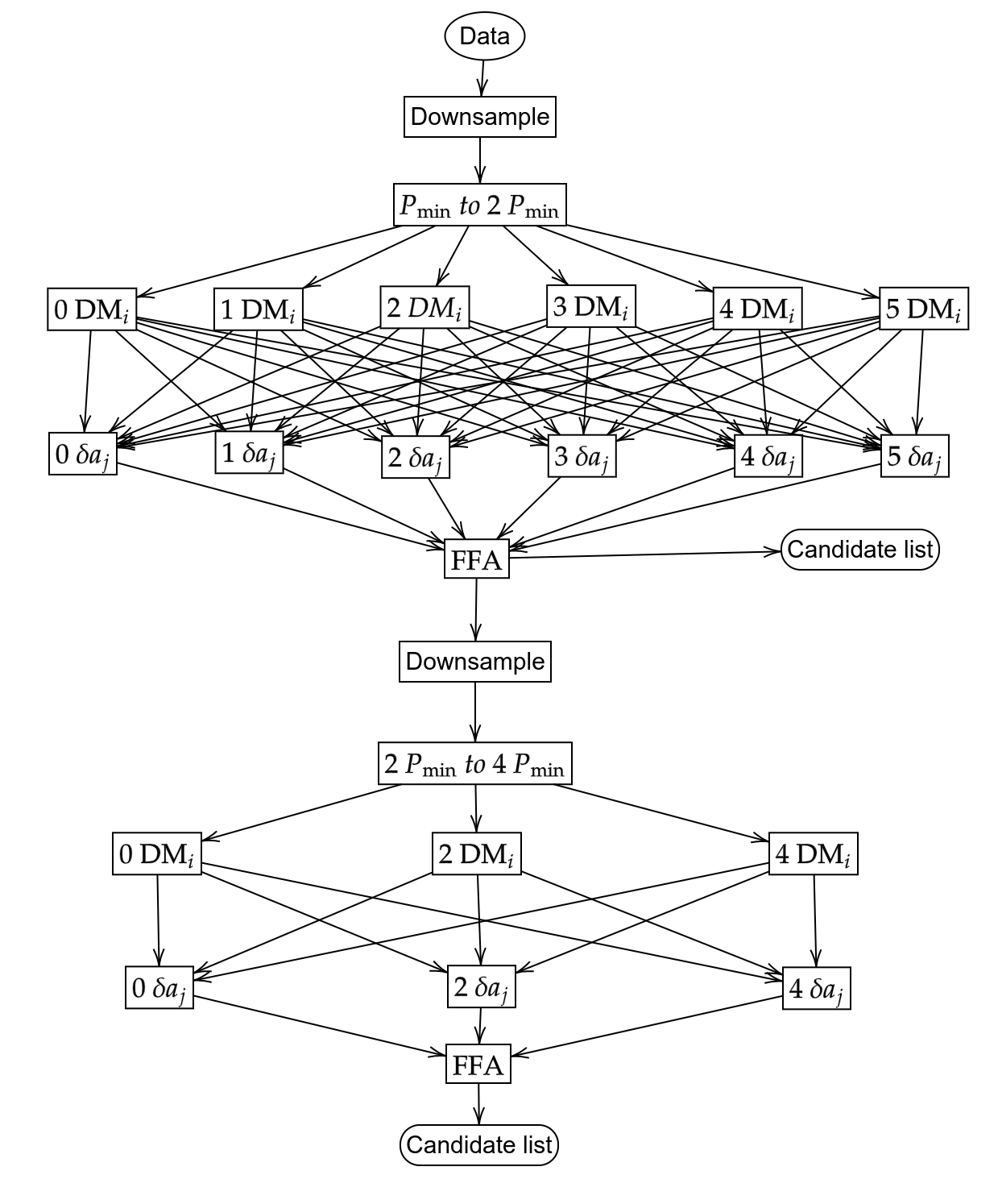}
 \caption{An example of the optimisation of the number of the dispersion and acceleration steps in the pipeline to search in the period range of $P_{\rm min}$ to $4 P_{\rm min}$, dispersion measure range of 0 to $5 \rm{DM}_i$, and the acceleration range of 0 to $5 \delta a_j$. This pipeline is separated into two period ranges. The first part of this pipeline is to search for the period range of $P_{\rm min} $ to $2 P_{\rm min}$ and to use the dispersion step of ${\rm DM}_i$ and the acceleration step of $\delta a_j$. In the second part, the period range is $2 P_{\rm min}$ to $4 P_{\rm min}$ with the dispersion step and the acceleration step being $2 {\rm DM}_i$ and $2 \delta a_j$, respectively. The existing time-series from the first part is reused in the second part of the pipeline.}
 \label{fig:steps}
\end{figure}

To correct for $\Delta t$, the time-series was resampled by shifting each data point in the time-series to match the expected arrival times for a range of trial accelerations. The acceleration step size is defined as the value that changes the arrival time by a bin over half the length of the observation $T_{\rm obs}$ (by resampling the data from the middle of the observation), which is written as 
\begin{equation}
\delta a = \frac{8  t_{\rm{samp}}  c}{T_{\rm{obs}}^2}.
\end{equation}
Consequently, the $j^{\rm th}$ acceleration step is written as 
\begin{equation}
\label{acc_step}
\delta a_j = (j-1) \frac{8  t_{\rm{samp}}  c}{T^2_{\rm{obs}}}.
\end{equation}
From the above equation, $\delta a \propto t_{\rm{samp}}$. As a result, the number of steps can be reduced (similarly to dispersion) as the data are downsampled. The relation between the downsampling factor, $n$, and the acceleration step is 
\begin{equation}
\delta a_{nj} = n \times \delta a_j.
\end{equation}
The key advantage of this optimisation is when the data are dedispersed at $\rm{DM}_i$ and resampled it at $a_j$, these data can be reused when the FFA reaches a fold at twice the period, as shown in Figure \ref{fig:steps}. As both of the DM and acceleration step are proportional to the minimum period searched, when the searched period range changes by a factor of two, the total number of acceleration and DM trials decreases by a factor of four.
As a result, this optimisation reduces the number of trials by approximately 86 per cent for the LowLat survey, as shown in Table \ref{tab:trials}.
\begin{table}
\caption{The comparison {of processing steps for a search} with and without the optimisation demonstrated in this work. $N_{r}$ is the total number of trials for each range. This optimisation reduced the number of trials for approximately 86 per cent for the selected period range.}
\label{tab:trials}
\resizebox{\columnwidth}{!}{
\begin{tabular}{c|ccc|ccc}
\hline
$P_{\rm min}$       & \multicolumn{3}{c|}{Without optimisation} & \multicolumn{3}{c}{With optimisation} \\ 
(s)     & $N_{\rm DM}$   & $N_{\rm acc}$   & $N_{ \rm r}$   & $N_{\rm DM}$  & $N_{\rm acc}$  & $N_{\rm r}$ \\ \hline
0.6     & 1,000 & 43 & 43,000  & 1,000 & 43 & 43,000 \\
1.2     & 1,000 & 43 & 43,000  & 500   & 22 & 11,000 \\
2.4     & 1,000 & 43 & 43,000  & 250   & 11 & 2,750  \\
4.8     & 1,000 & 43 & 43,000  & 125   & 6  & 750    \\
9.6     & 1,000 & 43 & 43,000  & 63    & 3  & 189    \\
19.2    & 1,000 & 43 & 43,000  & 32    & 2  & 64     \\
38.4    & 1,000 & 43 & 43,000  & 16    & 1  & 16     \\
76.8    & 1,000 & 43 & 43,000  & 8     & 1  & 8      \\
153.6   & 1,000 & 43 & 43,000  & 4     & 1  & 4      \\
307.2   & 1,000 & 43 & 43,000  & 2     & 1  & 2      \\ \hline
$N_{\rm total}$ &          &       & 430,000 &          &       & 57,783 \\ \hline
\end{tabular}
}
\end{table}

\subsection{The search pipeline}
\label{pipeline}
The initial stage of the search pipeline involves cleaning radio frequency interference (RFI) from the data. \texttt{RFIFIND} from \texttt{PRESTO}\footnote{https://github.com/scottransom/presto} was used to remove the brightest RFI bursts. The data were then dedispersed using the \texttt{PREPSUBBAND} routine to generate time-series at the specified DMs. The dedispersed time-series were resampled at each acceleration step using the \texttt{resample} routine in \texttt{SIGPYPROC}\footnote{https://github.com/ewanbarr/sigpyproc}. Afterwards, the RFFA was used to search in the shortest period range. For the next period range, the acceleration and dispersion steps doubled and half of the time-series generated previously were reused. This process was repeated to cover the whole period range.

Candidates with a S/N lower than a cut-off S/N were filtered out. The cut-off S/N was determined by calculating the false alarm rate (FAR) for the $N_{\rm{total}}$ number of trials. The FAR was calculated as follows:
\begin{equation}
\label{Equation:prob}
        \rm FAR= \rm Prob(>S/N)=\frac{1}{2} \left [1-erf \left (\frac{S/N}{\sqrt[]{2}}\right)\right].
        \end{equation}
The error function, {$\rm {erf}(x)=\frac{2}{\sqrt{\pi}} \int_0^x e^{-x^2} dx$}, can be solved numerically \citep[e.g.][]{press1992numerical}. Where the number of false detection for the search is calculated from FAR$\times N_{\rm total}$ where $N_{\rm{total}}$ was evaluated from 
\begin{equation}
    N_{\rm total}=N_{\rm FFA} \times  N_{\rm acc} \times  N_{\rm DM}.
\end{equation}
 \par $N_{\rm DM}$ and $N_{\rm acc}$ are shown in Table \ref{tab:trials}, and $N_{\rm FFA}$ was demonstrated by \citet{morello2020optimal} as 
\begin{equation}
    N_{\rm FFA}=\frac{T_{\rm obs}}{P_{\rm min}}N_{\rm bin}^2 \log{\frac{T_{\rm obs}}{P_{\rm min}}N_{\rm bin}^2}.
\end{equation}

Applying a limiting S/N to the data reduced the number of candidates greatly. However, the number of candidates observed per observation typically remains in the thousands which is impossible to inspect visually for a blind survey. In this work, we reduce the number of candidates using the kurtosis of the pulse profile's power distribution. Kurtosis is a property that quantifies the ``tailiness'' of a distribution; for a normal distribution, the kurtosis is 0. The kurtosis filtering has been typically used to detect bright RFI pulses in raw data ~\citep[for example][]{2007PASP..119..805N,2010MNRAS.406L..60N,2022MNRAS.510.1597P}. However, {here we use} the kurtosis score as a way to detect a pulsar-like signal but filtering out the low kurtosis profiles which were typically noise or weak RFI. This technique has a negative effect on the survey sensitivity to very broad pulse profiles. 
However, it is expected that such broad pulse profiles would have been identified in previous FFT based searches, as both FFT and FFA based pipelines should yield similar results for these types of profile, i.e. converging to predicted value from the radiometer equation \citep{morello2020optimal}.
The kurtosis limit was determined by comparing the kurtosis from a random noise profile to a profile containing a Gaussian. The comparison was conducted by simulating 2000 pulse profiles with pure noise and another 2000 Gaussian profiles with noise to represent pulsar pulse profiles. The duty cycle of these Gaussian profiles ranged from 1 per cent to 20 per cent. The kurtosis distribution from the simulation showed that the lowest kurtosis for the pulsar signal was approximately $-$0.5. We doubled this value to cover some extreme cases. As a result, only the candidates with kurtosis higher than $-$1.0 were inspected. Such criteria can reduce the number of candidates by 90 per cent.

\subsection{Search parameters}
\label{searchparams}
The data were searched with the aforementioned pipeline for a total period range of 0.6-s to 432-s. The longest searched period was chosen based on the assumption that the FFA requires at least 10 pulses to obtain more S/N than the single pulse searches \citep[see][for example ]{2010PhDT.......460K}. The shortest period search was limited by computational resources. As the number of search trials ($N_{\rm FFA}$) is proportional to $N_{\rm{samp}}^3$ for FFA searches \citep{morello2020optimal}, processing time increases quickly as the period reduces. We also chose the minimum search period to be 0.6-s, as this period covers a large portion of the known pulsars and still achievable with a reasonable processing time of order 72 h per beam with the available computing facility. The $N_{\rm{bins}}$ for folded profile was set to be 128 bins based on a duty cycle of $\sim1\%$, resulting in a $4.6875$-ms $t_{\rm{samp,1}}$ for the first period range according to Equation \ref{tsamp}.

For the maximum search DM, we used the YMW16  free electron distribution model \citep{yao2017new} to estimate the dispersion measure on a line-of-sight directly through the GC to the edge of the Galaxy. This extreme scenario gave a maximum DM of 3946 $\rm{pc}\,\rm{cm}^{-3}$. Thus, the DM range (DM$_r$) for this search\footnote{Note that NE2001 predicts a maximum DM for this line-of-sight to be 3396  $\rm{pc}\,\rm{cm}^{-3}$. } was set to be 4000 $\rm{pc}\,\rm{cm}^{-3}$.  The dispersion step size was calculated from Equation \ref{DM_step} with the HTRU-S LowLat's bandwidth (340 MHz) and central frequency (1352 MHz) with $i$ of 1 and $t_{\rm{samp,1}}$ of 4.6875 ms at 4.023  $\rm{pc}\,\rm{cm}^{-3}$. 

Because the GC is a dense environment, it is necessary for our acceleration range to be sufficiently wide to cover various kinds of binary companions. The acceleration range was chosen to be $\pm$ 128 m s$^{-2}$, which corresponds to a pulsar (of mass 1.4 M$_\odot$) with a companion up to the mass of a 37$M_\odot$, as e.g. black hole in a 12-hour orbital period \citep[see ][for calculations]{ng2015high}.

To improve clarity, the text was changed to: 
For the S/N limit, we used a S/N of 8.0. With this S/N, the number of false candidates is approximately 0.04 candidates per beam under the assumption that the data contains only white noise, which is exceptionally low.

To test the acceleration part of the pipeline, we generated 5600 artificial pulsars in various binary systems, using \textit{SIGPROC}'s \textit{FAKE} package \citep{2011ascl.soft07016L}. The simulated pulsars were selected to have spin periods randomly chosen between 1.0 to 6.0-s, with pulse duty cycles ranging from 1 to 25 per cent.\footnote{We chose this period range rather than the full period range (0.6-s to 430-s) because the RFFA downsamples the time-series to the optimal time resolution, making the period range arbitrary.} The companion mass was randomly selected between 0 and 37 $M_{\odot}$ with a fixed orbital period of 12 h at an orbital phase of 0.25, making it the easiest orbital phase to detect for an acceleration search. We then also generated the same pulsars without the binary companions as an isolated realisation. We compared the S/N values resulting from the pipeline from the isolated and accelerated realisations to determine the loss S/N. Our simulations showed that 90 per cent of the acceleration search FFA results have S/N greater than 95 per cent of the S/N detected from identical isolated pulsars.

\section{Results}
\label{Results}
\subsection{Redetection of known pulsars}
\label{known}
Of the ten previously known pulsars in the targeted region, two of them are millisecond pulsars, which are outside of our searched period range (PSR~J1747$-$2809 and PSR~J1745$-$2912). Three of them (PSR~J1745$-$2758, PSR~J1747$-$2802, and PSR~J1750$-$28) were detected with previous FFT based processing  \citep{ng2015high,cameron2020high}. We also detected PSR~J1746$-$2856 that was not reported in the previous FFT based survey processing. The GC magnetar PSR~J1745$-$2900, and three other long period pulsars (PSR~J1746$-$2850, PSR~J1746$-$2849, and PSR~J1746$-$2856) were not detected with the FFT or FFA pipelines, while PSR~J1745$-$2910 had never been detected at this observational frequency before. The details about eight pulsars inside our search period range are shown in Table ~\ref{table:known}. 

The reasons for the non-detections are as follows: PSR J1745$-$2900 and PSR J1746$-$2850 are known to be transient pulsars and it is possible that these pulsars were not active during our observations. We further folded those observations containing these pulsars using the ephemeris from \textsc{psrcat} and found no pulsations. PSR J1745$-$2910 has never been detected afterward \citep[see e.g.][]{macquart2010high,eatough2021multi}, suggesting that it could also be a transient pulsar. Meanwhile, PSR J1746$-$2849 exhibits a substantial scattering tail at the L-band \citep[266 ms][]{deneva_2010} and has a faint average flux, leading to an anticipated low S/N. When we folded this pulsar at the closest pointing using the current ephemeris reported in PSRcat, we did not find any pulsations with an S/N greater than 5.

The re-detection of PSR J1746$-$2856 with a relatively high S/N suggests that it was overlooked due to book keeping error in the earlier FFT-based processing. This assumption is reinforced by the detection of this pulsar during the reprocessing of the HTRU-S low-lat using an FFT-based GPU-accelerated pipeline (Sengar et. al., In prep.).

Comparing the S/N from both the FFA and FFT-based pipelines demonstrates that the S/N from the FFA is consistently higher than that from the FFT, as predicted by \citet{morello2020optimal}.

\begin{table}
\caption{Known slow pulsars in the HTRU-S LowLat using the FFA pipeline. {Note that} PSR J1745$-$2758 was detected at the 2$^{\rm{nd}}$ harmonic of the period (0.975s). }
\label{table:known}
\resizebox{\columnwidth}{!}{
\begin{tabular}{llllll}
\hline
NAME       & P0       & Flux density at 1400 MHz & duty cycle $^{\diamondsuit}$& $S/N_{\rm FFA}$ & $S/N_{\rm FFT}$ \\
           & (s)      & (mJy) & (\%)      &             &             \\ \hline
J1745$-$2758 & 0.487528 & 0.15  & 6.1     & 8.8         & 8.4         \\
J1745$-$2900 $^{\spadesuit}$ $^{\heartsuit}$& 3.763733 & 0.9$^\clubsuit$     & 8.3      & -           & -           \\
J1745$-$2910 $^{\spadesuit}$ & 0.982    & -     & 8.0     & -           & -           \\
J1746$-$2849  $^{\spadesuit}$& 1.47848  & 0.4   & 8.2     & -           & -           \\
J1746$-$2850  $^{\spadesuit}$ $^{\heartsuit}$ & 1.077101 & 0.8   & 5.6     & -           & -           \\
J1746$-$2856  $^{\spadesuit}$ & 0.945224 & 0.4     & 4.8     & 14          & -           \\
J1747$-$2802 & 2.780079 & 0.5   & 1.0     & 15.6        & 14.6        \\
J1750$-$28   & 1.300513 & 0.09  & 1.3     & 12.8        & 8.7         \\ \hline
\end{tabular}}
\small $^{\spadesuit}$Pulsars that are not detected in ~\cite{ng2015high}.\\
\small $^{\heartsuit}$Pulsars that show{s} high flux variation.\\
\small $^{\diamondsuit}$ The duty cycles were calculated from $\frac{W50}{P0}${,} using data from \textsc{PSRcat} \citep{2005AJ....129.1993M}\footnote{https://www.atnf.csiro.au/research/pulsar/psrcat/}. \\
\small $^{\clubsuit}$ For this work, we estimate the L-band flux density from extrapolating the flux and spectral index reported in \citet{2017MNRAS.465..242T}.
\end{table}

\subsection{PSR J1746$-$2829: A new discovery}
\label{New pulsar}
A new pulsar, PSR~J1746$-$2829 was found during the reprocessing of LowLat data with an FFA S/N (S/N$_{\rm FFA}$) of 11.2. Although it is not directly in the central parsec of the GC, its proximity ($\sim$ 0.5$^\circ$) makes it an interesting source for comparison with the known pulsars in this region. We found that the pulsar has a high DM ($1309~{\rm pc~cm}^{-3}$) and a long period ($1.89$s), full parameters of the parameters for this pulsar can be found in Table \ref{tab:my-table}. This pulsar was also found in the later reprocessing of the HTRU-S low-lat using FFT based GPU-accelerated pipeline (Sengar et. al., In prep.) with  S/N$_{\rm FFT}$ of 8.6, $\sim$20 per cent less than S/N$_{\rm FFA}$.

The reason this pulsar was overlooked in the previous FFT survey is due to the use of 2-bit digitisation in the decimation code. A bug was found in the code that effectively sampled the data at 1-bit, leading to a 
 approximately 25 per cent loss in sensitivity. This pushed the pulsar below the 8 sigma threshold for folding in the older FFT pipeline. This part was removed in the recent FFT reprocessing (Sengar et. al., In prep.) then the pulsar became detectable by the FFT \citep[see][for more details]{sengarSearchingRadioPulsars2023}.

\begin{table}
\caption{Timing parameters for newly discovered pulsar with 1-$\sigma$ uncertainty represented in parentheses.}
\label{tab:my-table}
\begin{tabular}{ll}
\hline
Parameter                            &               \\ \hline
Name                                 & J1746-2829    \\
Right ascension (J2000)$^{\heartsuit}$               & 17$^{\rm h}$46$^{\rm m}$15$^{\rm s}$(14)      \\
Declination (J2000)$^{\heartsuit}$                   & -28{\degr} 29{\arcmin} 32{\arcsec}(38)     \\
Galactic latitude, b ($^\circ$)              &  0.117(60)             \\
Galactic longitude, l ($^\circ$)                & 0.445(56)              \\
Spin period  (s)                     & 1.888928609337(9) \\
Period derivative (s s$^{-1}$)$^{\diamondsuit}$             & 1300(30) $\times  10^{-15}$  \\

Epoch of period                      & 58564.0 \\
Dispersion measure (cm$^{-3}$ pc)      & 1309(2)  \\
Estimated distance$^{\spadesuit}$ (kpc)      & 8.2 \\
Rotation measure (rad m$^{-2}$)         & $-$743(14)      \\
Scattering time (ms)                 & 67(3)          \\
Average flux density at 1400 MHz (mJy) & 0.55(6)       \\
Inferred $B_{\rm field}$ (G) & 5.0 $\times 10^{13}$       \\
Inferred characteristic age (yr) & 23 $\times 10^{3}$       \\
Spin-down luminosity (erg s$^{-1}$)       & 8.4 $\times  10^{33}$              \\
      \\
Flux density spectral index & $-$0.9(1)     \\ 
Start MJD   & 58564 \\
Finish MJD  & 59838 \\ \hline

\end{tabular}
\\
\small $^{\spadesuit}$ Estimated with \citet{yao2017new}\\
\small $^{\heartsuit}$ Position obtained from the MeerKAT's detections. \\
\small $^{\diamondsuit}$ This uncertainty considers a contribution from positional uncertainty.
\end{table}

\subsubsection{Follow-up observations}
After its discovery, PSR J1746$-$2829 was observed with the Parkes telescope using the 21-cm Multibeam receiver for seven epochs from April to July 2018. However, even the 72{-}min observations at Parkes were yielding S/N of only approximately 8-9, so observations were also made with the 100-m Effelsberg telescope. While the larger diameter of the Effelsberg telescope leads to higher sensitivity, the beam size {($\theta_{ \rm b}$)} is smaller at the same frequency. Hence, the Effelsberg's 21-cm receiver\footnote{This receiver has an effective bandwidth of 250 MHz see https://eff100mwiki.mpifr-bonn.mpg.de} was used to make 1800-s `gridding' observations \cite[see][for more details]{cruces2021fast}. This technique is performed by observing the pulsar multiple times with a small offset from the central beam to improve the position, narrowing down the uncertainty to within the width of the Effelsberg beam
at 21-cm (0.163$^{\circ}$). The pulsar was subsequently observed for seven epochs with the same receiver and observation time using the Effelsberg telescope.

The installation of the Ultra-Wide Band (UWL) \citep{hobbs2020ultra} receiver at the Parkes telescope provided a substantially larger bandwidth of $\sim$ 3 GHz. The pulsar was {therefore also} observed for {14} epochs from March 2019 to October 2022 and was detected at the upper frequency range of the UWL receiver at $\sim$ 4 GHz, implying that this pulsar might have a flat spectrum (see Section \ref{params} for details).

In order to reduce the positional uncertainty of the pulsar further, observations were scheduled with the MeerKAT interferometer \citep{2016mks..confE...1J}. The long baselines between individual dishes reduce the size of the synthesised beam, thus giving an improved localisation of the source of interest. The FBFUSE ~\citep{2018IAUS..337..175B} system offers the capability of beamforming \citep{2021JAI....1050013C} and recording up to 864 tied array beams and producing SIGPROC format filterbank data. Using the UWL position of PSR J1746$-$2829 as a reference, 480 beams were tiled around the position with an integration time of 9 min at 1.28 GHz. This covered roughly a 25 arcmin radius. Beams within the UWL positional uncertainty were folded/searched with the pulsar parameters. The only detection was obtained in a beam centred at RA 17$^{\rm h}$46$^{\rm m}$15$^{\rm s}$.04 and DEC -28$\degr$29$\arcmin$32$\arcsec$.40. Since the synthesised beams were elliptical, the positional uncertainties in the major and minor axis were 50 and 80 arcsec respectively.

\begin{figure}
 \includegraphics[width=\columnwidth]{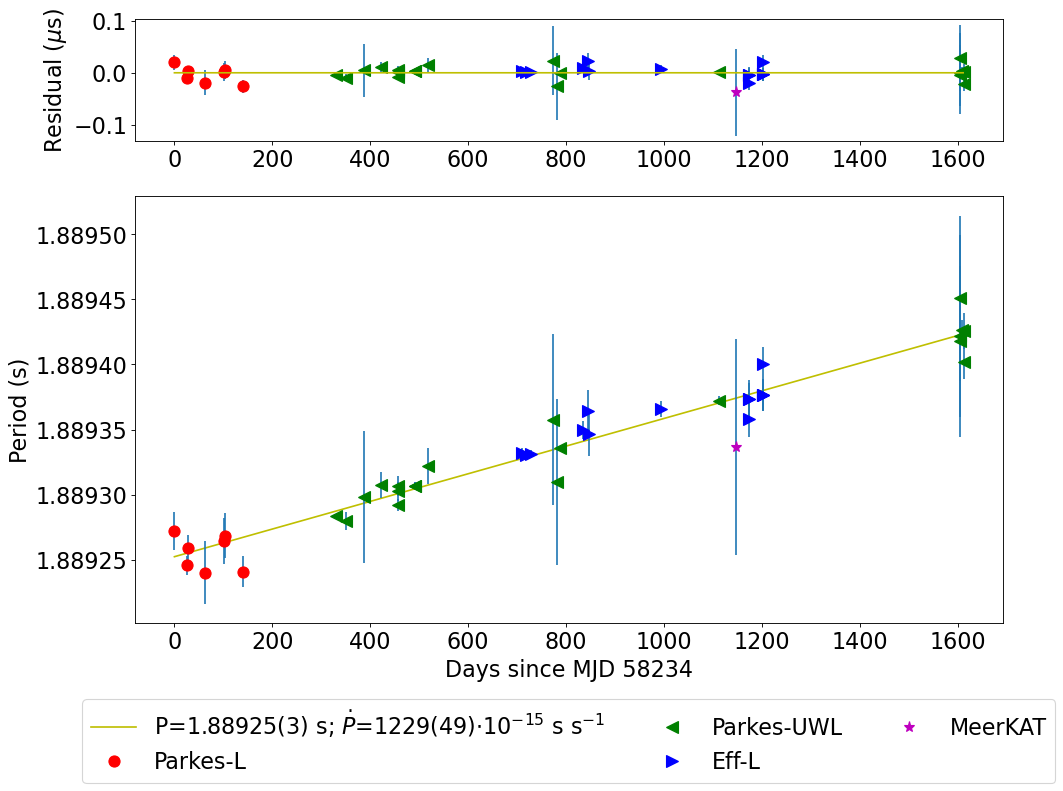}
 \caption{Barycentric spin period evolution of PSR J1746$-$2829. The spin period of this pulsar can be described by a simple linear function ($P(t)=\dot{P}t+P_0$).{The different colours represent the different instruments used.} }
 \label{fig:period-mjd}
\end{figure}

To estimate the spin-down rate of the pulsar and derive an initial timing ephemeris, we modeled the measured spin period in each observation as a first-order polynomial,

 as shown in Figure \ref{fig:period-mjd}. Consequently, the pulsar's spin period evolution is dominated by a period derivative ($\dot{P}$) which corresponds to 1229(49)$\times  10^{-15}$s s$^{-1}$. This initial timing solution was then used to fold all of the observations of this pulsar.

The phase-connected timing solution (where every rotation of the pulsar is modelled from the first observation to the last) was only possible with the UWL observations from MJD 58564 to 59838 due to the combination of a large timing gap and highly RFI contaminated observations. This timing solution is shown in Table \ref{tab:my-table}. Although the timing solution was not fully phase-connected to the other data sets, it confirms the high $\dot{P}$ of this pulsar. The position was not fitted because it yielded a location outside the MeerKAT beamwidth, suggesting that the current timing position uncertainty is still larger than the position uncertainty derived from the MeerKAT pulsar search observation.   Assuming this pulsar to be a magnetic dipole radiator with canonical neutron star mass ($1.4 M_\odot$) and radius ($10$ km), the surface magnetic field ($B_{\rm field}$) can be estimated: $B_{\rm field} \sim 5 \times  10^{13}$ G with a characteristic age of $\sim$ 23000 yr. The second-period derivative was found to be highly correlated with the position uncertainty, which makes it less likely to be intrinsic to the pulsar.

To further constrain the position, we modelled the MeerKAT beam as a Gaussian function (see Equation \ref{eq:offset1}). Since this pulsar was detected with an S/N of 12 in only one of the tile-array beams of MeerKAT, and considering that the S/N limit for the MeerKAT searches was 7, we used the beam to determine how far the pulsar could be located without being detected in the neighbouring beam with an S/N < 7 \citep[see e.g.][]{2023arXiv230210107J}. The beam and the current position uncertainty are shown in Figure \ref{fig:MeerKATimage}.

\subsubsection{Polarisation, spectral distribution, pulse profile evolution, and scattering time}
\label{params}
The synthesised beam width of the MeerKAT telescope ($\sim$ 1 arcmin) is approximately six times smaller than that of the upper band at UWL receiver ($\sim$ 6.6 arcmin) \citep{hobbs2020ultra}. As a result, we used the beam position from the MeerKAT observations as this pulsar's position, minimising the impact of a potential position offset on the measured flux density. Now that the angular offset had less effect on the flux of the pulsar, the intrinsic spectrum could be measured. We carried out an observation with UWL receiver to measure the polarization properties, the spectral index, the profile evolution, and the scattering time. Polarisation calibration was done by observing a noise diode at 45$^{\circ}$ to the receiver dipoles for 90–120s prior to the observation of the pulsar. The data were calibrated for flux using on- and off-source scans of radio sources with known, stable flux densities, e.g. the radio galaxy Hydra A\citep[obtained from,][]{2020PASA...37...20K}.

\begin{figure} 

 \includegraphics[width=1.0\columnwidth]{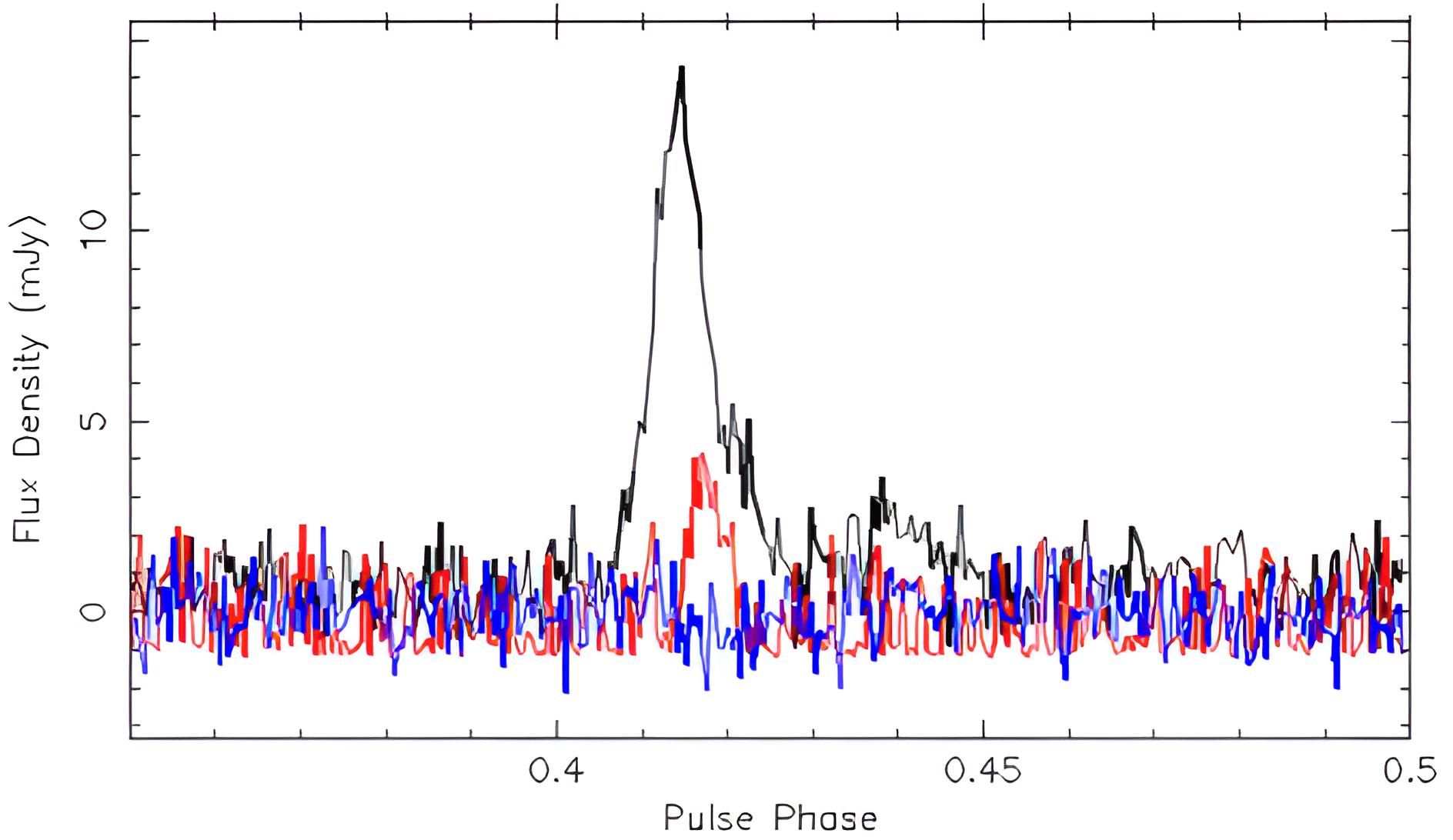}
 \caption{Folded pulse profile of PSR J1746-2829 at 3.2GHz showing a narrow pulse profile with a low linear (Red) polarization fraction and no significant circular (Blue) polarisation after correcting for RM.}
 \label{fig:lum}
\end{figure}

We used the \textit{RMcalc} code \citep{porayko2019testing} to measure the rotation measure (RM). This approach was based on~\cite{brentjens2005faraday} RM synthesis method and presented the estimates using the Bayesian Generalized Lomb-Scargle Periodogram (BGLSP) technique. For consistency, we also used the \textit{RMFIT} routine from  \textit{PSRCHIVE} \citep{2012AR&T....9..237V}, which is based on an optimisation of the linear polarization fraction. The searched RM range was set according to the analysis by \cite{schnitzeler2015rotation}, resulting in the RM range of $\pm$78481 rad m$^{-2}$. \textit{RMcalc} gave a rotation measure of -743 $\pm$ 14 rad m$^{-2}$ while \textit{RMFIT} gave $-$797 $\pm$ 39 rad m$^{-2}$. As the results are consistent, the RM from \textit{RMcalc} with a lower uncertainty has been used. After applying this RM we detected linear polarisation  of $\sim$ 20 per cent without significant circular polarisation. The RM of this pulsar, which is notably high i.e. $\geq$ 500 rad m$^{-2}$, is consistent with the other known pulsars the direction of the GC \citep{2016MNRAS.459.3005S,2023MNRAS.524.2966A}. The pulse profile with polarisation is shown in Figure \ref{fig:lum}.

\begin{figure}
 \includegraphics[width=\columnwidth]{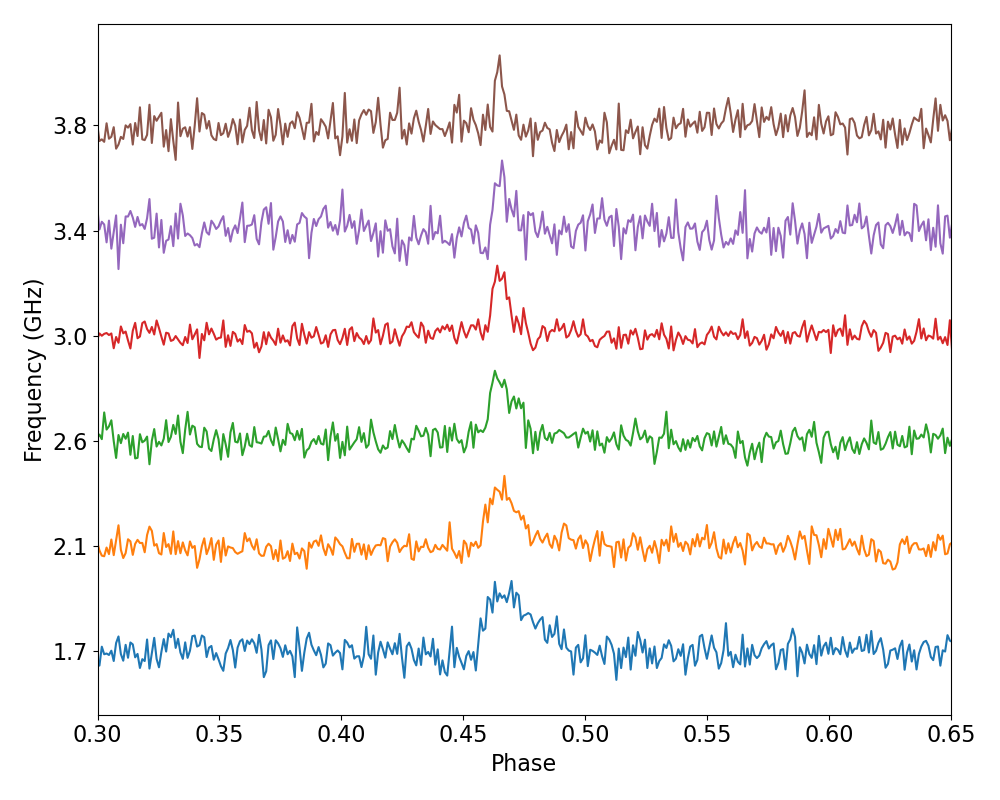}
 \caption{Frequency vs pulse-phase plot for PSR J1746$-$2829 from an 82-min observation with the UWL receiver, containing 6 cleaned channels and 1024 pulse phase bins.}
 \label{fig:freq}
\end{figure}

{Subsequently, the observation was split into eight subbands to study the pulse profile evolution over frequency.} The pulse profile at the uppermost band (3.84 GHz) showed a profile that could be described by a simple Gaussian function\footnote{Due to its low significance, we did not consider the additional structure near phase 0.44 in the pulse profile, as it origin, whether intrinsic to the pulsar or residual RFI, remains unclear.} with the full width at half maximum (FWHM) of 10.15$\pm$0.07 ms, resulting in a small duty cycle for this pulsar of 0.537 $\pm$ 0.04 per cent. The pulse profile shows an increasingly broader tail towards lower frequencies, as shown in Figure \ref{fig:freq}. 

Assuming no pulse profile evolution over frequencies, the broadening time is measured by fitting for an exponential decay\footnote{$e^{-\frac{t}{t_{ts,f}}}$}  with a characteristic time defined as:
\begin{equation}
    \tau_{\rm{ts,f}}=\tau_{\rm{ts}} (\frac{f}{1000 \, \rm{MHz}})^{\alpha_{\rm{sc}}}
\end{equation}
where the reference frequency is 1000 MHz. 
The frequency-phase pulse profile was modelled as a Gaussian pulse convolved with exponential decay using \texttt{Pulse Portraiture} \citep{2014ApJ...790...93P,2019ApJ...871...34P}, $\tau_{\rm ts}$ was determined through least square minimisation, with $\alpha_{\rm sc}$ at $-$4.0.\footnote{Depending on the screen and DM.} The resulting scattering time for this pulsar at 1000 MHz as 67 $\pm$ 3 ms.

The scattering time for this pulsar is notably lower than that of other pulsars in the GC, with $\tau_{\rm sc} \geq 200$ ms at 1.4 GHz \citep{johnston2006discovery,deneva2009discovery}. This is expected given its 0.5$^\circ$ offset from the GC. However, the $\tau_{\rm sc}$ for this pulsar is approximately 4 ms at 2.0 GHz. Such a scattering time could potentially smear out the pulsations from some fast recycled pulsars, even at a 0.5$^\circ$ separation.

\subsubsection{{Radio flux density and spectrum}}
\label{specflux}
{To study the spectrum, we choose observations with $T_{\rm obs}$ longer than one hour with the UWL receiver, resulting in two observations at MJD 59022 and MJD 59491. The observation at MJD 59091 was made at the current best position. However, the observation from MJD 59022 was made before the current position was determined. It has a position offset ($\theta$) of $\sim$ 2 arcmin. An offset $\theta$ is causing the observed flux  density to be reduced as } 

  \begin{equation}
  \label{eq:offset1}
      S_{\rm obs}=S_{\rm f}e^{\frac{-\theta^2}{2 \sigma^2}},
  \end{equation}
  {where $S_{\rm obs}$ is the observed flux density. This equation assumes that the telescope response pattern is a Gaussian, where $\sigma$ is a Gaussian rms width calculated from 
  \begin{equation}
  \label{eq:offset2}
  \sigma=\frac{\theta_{\rm b}}{2\sqrt{2 \ln{2}}},
  \end{equation} 
 using $\theta_{\rm b}$ in each frequency band as published in \cite{hobbs2020ultra}.}
\par { The $S_{\rm obs}$ was determined using the \textit{PSRFLUX} routine from \textit{PSRCHIVE} \citep{2012AR&T....9..237V} with the 2D pulse profile template derived from the profile discussed in the previous section. After we compensated for the flux density reduction due to the offset using Equation \ref{eq:offset1} and Equation \ref{eq:offset2}, the spectrum was modeled as a power law with a spectral index ($\alpha$),  
\begin{equation}
\label{equation:index}
	S_{\rm f}=S_{f_{\rm{ref}}}(\frac{f}{f_{\rm{ref}}})^{\alpha},
\end{equation}
where $f_{\rm{ref}}$ is the reference frequency which is 1400 MHz.  We measure the spectral index to be $-$0.8 $\pm$ 0.1 and $-$0.9 $\pm$ 0.2 with a flux density at 1400 MHz of 0.48 $\pm$ 0.01 mJy and 0.35 $\pm$ 0.01 mJy for MJD 59022 and MJD 59491 respectively, confirming that this pulsar has a relatively flat spectrum compared to the typical pulsar population, which is in range of $-$1.4 to $-$1.6 ~\citep{spectral0,spectral} and showing no sign of spectral index variation overtime, which has been found in two radio loud magnetars  \citep{iram,2020MNRAS.498.6044C}.}

 To explore radio flux distribution and evolution in time, the radio light curve was calculated using two methods. First, all L-band flux densities were estimated from the radiometer equation using the effective bandwidth, observation time and system equivalent flux density for Parkes \citep{keith2010high}, Effelsberg\footnote{https://eff100mwiki.mpifr-bonn.mpg.de} and MeerKAT \citep{2020PASA...37...28B}. In addition, all of the observations before the MeerKAT observation at MJD 59381 had a positional offset which decreases the flux density which can be corrected using Equation \ref{eq:offset1}. The flux density from the MeerKAT observation is still affected by the position offset, which is representing as a large asymmetric uncertainty, where the upper limit represents the scenario where the pulsar is located at the edge of the uncertainty. Moreover, we also search for a point source in the GC image mosaic from MeerKAT \citep{2022arXiv220110541H} (see Section \ref{discussion}). Only one point source was found within the positional uncertainty, hence the flux density of this source (0.36 $\pm$ 0.04 mJy) was used as an upper limit for epochs that covered this source (MJD 58281,58286 and 58287).

  {Secondly, since more than 75 per cent of the UWL observations at approximately 1400 MHz were heavily polluted by RFI, the channels around 1400 MHz were removed. In this case, the flux density at 1400 MHz is determined by the flux density from the upper frequency data and the spectral index from all available UWL observations with polarisation and flux calibrators, using the average spectral index, $\alpha=-0.9(1)$. }
 \par
\begin{figure}
 \includegraphics[width=\columnwidth]{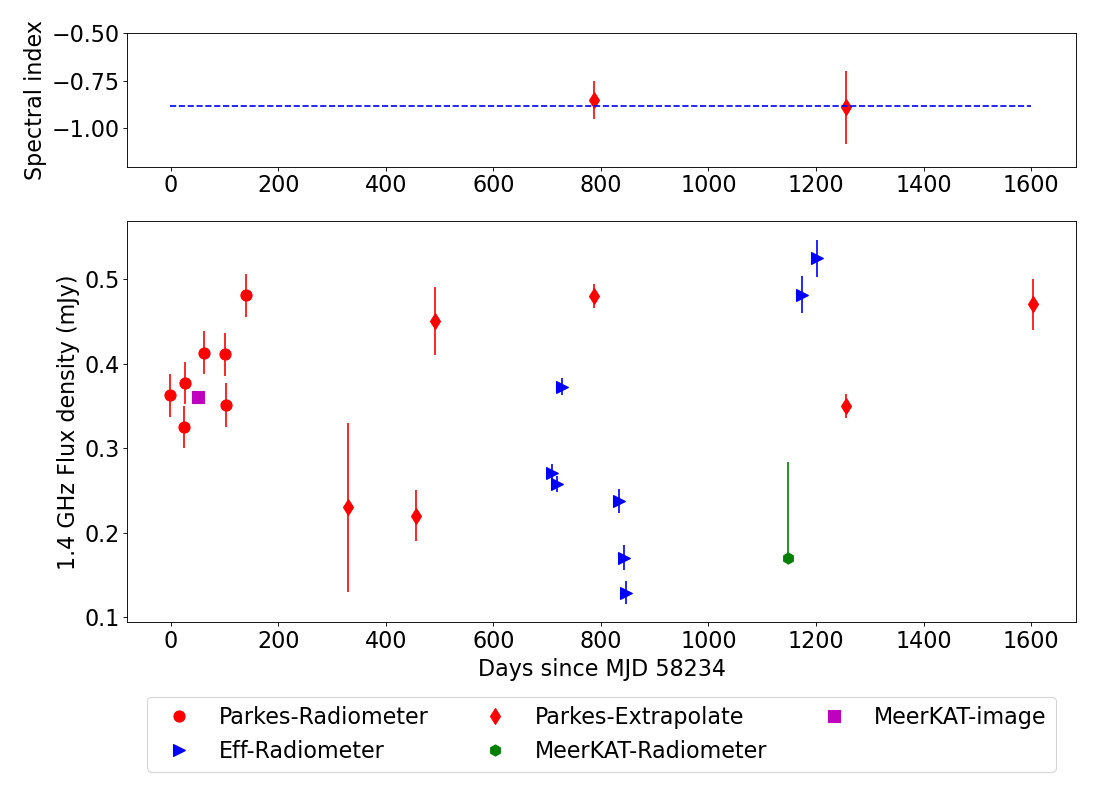}
 \caption{{Upper panel: Spectral indexes ($\alpha$) at two epochs showing no significant $\alpha$ variation, the dashed line represents the average $\alpha$. Lower panel: Radio light curve of PSR J1746$-$2829. Red circles represent the flux densities from Parkes L-band observations while the blue triangles represent the L-band fluxes densities from the Effelsberg telescope. The green hexagon represents the flux density calculated using the radiometer equation based on the MeerKAT follow-up observation of this pulsar. The uncertainty here also includes the contribution from any position offset. The magenta square represents the flux density derived from the only source within the position uncertainty (see Section \ref{discussion}). The red diamonds show the 1400 MHz flux densities extrapolated from the UWL observations assuming the spectral index of -0.9. } }
 \label{fig:flux_mjd}
\end{figure}

   The flux density is plotted against epoch of observation in Figure \ref{fig:flux_mjd}. The radio light curve shows large flux density variations (more than 50 per cent), corresponding to a characteristic of the high $B_{\rm field}$ pulsars and magnetars \citep[see e.g.,][]{2017MNRAS.468.1486D}. According to NE2001 \citep{2002astro.ph..7156C}, the estimated diffractive scintillation time scale ($t_{\rm d}$) at this location is approximately 1.3 s at 1000 MHz. This is significantly shorter than the observation time. NE2001 also predicted the scintillation bandwidth ($\Delta f_{\rm d}$) to be $3 \times 10^{-6}$ MHz, \citet{1990ApJ...352..207S} demonstrated that the reflective scintillation time scale ($t_{\rm r}$) can be estimated from 
  \begin{equation}
      t_{\rm r}=\frac{4f_{\rm obs} t_{\rm d}}{\pi \Delta f_{\rm d}}.
  \end{equation}
  The $t_{\rm r}$ at $f_{\rm obs}$ of 1000 MHz is $\sim 20$ yr, which is significantly longer than the flux density variation timescale ($\sim$ months) as shown in Figure \ref{fig:flux_mjd}. As a result, the observed flux density variation is unlikely dominated by the interstellar scintillation.
  
\subsection{X-RAY DATA}

{ Given the similarities between PSR J1746-2829 and some magnetars (see Section \ref{sec:mag}), we have considered the possibility of detecting the new pulsar in X-rays, given that magnetars are often detectable at these energies. We searched for X-ray emission  by cross-referencing with the sources from GC Chandra X-ray survey. One source was found with an angular separation of 0.441 arcmin from the centre of the MeerKAT radio beam where this pulsar was detected \citep[][]{Muno_2009}. The location of this source at the edge of the tile array beam makes it unlikely to be the pulsar as it was not detected in any neighbouring beams.}

An upper limit for X-ray emission was estimated from the processed and cleaned Chandra composite images\footnote{Obs ID:07045,07044,02294,18326,18329} that covered the whole positional uncertainty. We searched for the X-ray pixel with the highest count that is not within one pixel from the known X-ray source and used that as the upper limit for the number of photons, which was then converted to a count rate by dividing by the exposure time. \textit{WebPIMMS}\footnote{https://heasarc.gsfc.nasa.gov/cgi-bin/Tools/w3pimms/w3pimms.pl} was used to convert from the Chandra telescope count rate to X-ray luminosity using the spectrum (blackbody with kT = 1 keV) of the GC magnetar PSR~J1745$-$2900 \citep{2013ApJ...775L..34R} and the column density $N_H$ at the pulsar's position\footnote{https://heasarc.gsfc.nasa.gov/cgi-bin/Tools/w3nh/w3nh.pl} $N_H=9.42 \times  10^{21} \rm{cm}^{-2}$. The upper limit for X-ray flux for each epoch shown in Table~\ref{tab:X-ray}, is approximately $10^{32}$ to $10^{33}$ erg/s, which is 100 times less than the spin-down luminosity, $\sim 10^{34}$ erg/s. In terms of higher-energy emission, specifically gamma-rays, no point source has been reported within 1 arcmin of this source \citep{2022ApJS..260...53A}. This absence is expected not only due to the considerable distance of the pulsar but also because the region exhibits a high gamma-ray background level, reducing the number of detectable point sources \citep[see, e.g.,]{2019ApJ...871...78S}.

\begin{table}
\caption{{Results from X-ray archival images. All of them were observed before the start of the timing campaign and showing the minimum X-ray flux of $10^{32}$ to $10^{33}$ erg s$^{-1}$ which is approximately 1 to 10 per cent of this pulsar's spin down luminosity. The count rate uncertainties were calculated from root-mean-square of the count rate in the MeerKAT beam.}}
\label{tab:X-ray}
\begin{tabular}{llll}
\hline
MJD      & Count rate (photons ks$^{-1}$) & $L_x 10^{32}$ (erg s$^{-1}$) & $T_{\rm obs}$ (ks) \\ \hline
54154.30 & 0.135(14)                   & 1.05(11)                     & 37             \\
54040.59 & 0.132(15)                   & 1.02(12)                     & 38             \\
52107.89 & 0.258(20)                   & 2.00(16)                     & 12             \\
57584.33 & 1.036(69)                   & 8.04(54)                     & 1.9            \\
57584.43 & 1.036(65)                   & 8.05(50)                     & 1.9            \\ \hline
\end{tabular}
\end{table}
\section{Discussion}
\label{discussion}

This evidence of increased sensitivity from the FFA based pipeline compared to the FFT based pipeline can be seen in the S/N of the detected pulsars in Table \ref{table:known}, as the FFA always has a higher S/N than in the FFT. This reflects the previous results reported in  \citet{cameron2017investigation,parent2018implementation}. However, it should be noted that our FFA pipeline has reduced sensitivity to pulsars with duty cycles of less than $\sim$ 1 per cent, due to the number of bins used in the fold.

The newly discovered pulsar, PSR~J1746$-$2829, has a flat spectrum, long period, and a relatively high $\dot{P}$. 
These are characteristics that are similar to those of magnetars. However, the degree of linear polarisation is surprisingly low if it is indeed a magnetar (see discussions below). The parameters of this pulsar are shown in Table \ref{tab:my-table}.

Recently, \citet{2022arXiv220110541H} published a deep radio image of the GC made with the MeerKAT telescope at 1.28 GHz. The mosaic image also covered the location of PSR J1746$-$2829. The expected flux density of PSR J1746$-$2829 at the observing frequency of the imaging survey was calculated using Equation \ref{equation:index} with $S_{\rm{1400}}$ and $\alpha$ reported in Section \ref{specflux}, resulting in an expected flux density of $0.57^{+0.098}_{-0.070}$ mJy.

\begin{figure*}
 \includegraphics[width=0.8\textwidth]{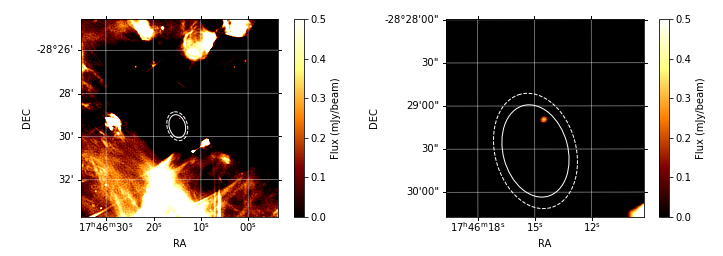}
 \caption{The L-band observations image mosaic obtained at MeerKAT \citep{2022arXiv220110541H}. Left: The MeerKAT pulsar search beam of PSR J1746$-$2829 (white dashed ellipse) reveals a point source with 0.36 mJy flux density and 0.04 mJy rms in the field with a position uncertainty derived from non-detections in the other beams from the follow-up observation with MeerKAT (white ellipse). Right: The magnified version of the image on the left, displaying the location of the point source in the MeerKAT pulsar search beam.}
 \label{fig:MeerKATimage}
\end{figure*}

As shown in Figure \ref{fig:MeerKATimage}, a point source was detected with a peak flux density of 0.3 mJy visible within the position uncertainty of the pulsar. The sensitivity of the imaging survey \citep{2022arXiv220110541H} implies that the pulsar should be visible, and the fact that there is only one point source within the positional uncertainty and the flux density of this pulsar varies between 0.1 mJy to 0.6 mJy suggests that it may indeed be the pulsar. Unfortunately, neither the spectral index nor the polarisation of this point source is reported, which would help to further confirm the association. 

Using a kick velocity of 380 km s$^{-1}$ reported by \citet{2006ApJ...643..332F} and assuming that this pulsar is located 8.2 kpc from Earth, the upper limit for the angular separation from the birthplace of this pulsar is approximately 4 arcminutes. According to the MeerKAT L-band Mosaic at the GC\citep{2022arXiv220110541H}, there are several sources that could be supernova remnants within 4 arcminutes from this pulsar. Consequently, a further proper motion measurement is required to determine if this pulsar is associated with any of these sources.

\subsection{Is PSR J1746$-$2829 a magnetar?}
\label{sec:mag}
Magnetars are a group of neutron stars characterised by an implied surface magnetic field strength orders of magnitude larger than that of the normal pulsar population. 
They are also found to have relatively long spin periods ($\geq$ 1 s) \citep[e.g.,][]{2017ARA&A..55..261K}.
Magnetars predominately emit X-rays and gamma rays, sometimes with a luminosity that exceeds the spin-down luminosity for a rotating magnetic dipole ($\dot{E}=4\pi^2 I \dot{P}/P^3$). For this reason, magnetar emission is believed to be powered by the decay of the magnetic field energy, rather than purely by rotation. Most magnetars are radio-quiet; out of the 31 magnetars discovered to-date \footnote{\url{http://www.physics.mcgill.ca/~pulsar/magnetar/main.html}} \citep{2014ApJS..212....6O}, only six of them are detectable at radio frequencies \citep{2006Natur.442..892C,2007ApJ...666L..93C,2010ApJ...721L..33L,2011ApJ...730...66L,2012MNRAS.422.2489L,2013Natur.501..391E,2013MNRAS.435L..29S,2020ATel13553....1K,2020ApJ...896L..37L}. The radio-loud magnetars are usually observed to be transient in nature, some magnetars have been known to be in a quiet state for years before becoming active \citep[e.g.][]{2013ATel.5020....1M,2018ATel12284....1L}. Most of these radio loud magnetars show flat radio spectra ($\alpha \leq -1.0$) \citep{
2007MNRAS.377..107K,
2007ApJ...666L..93C,
2013MNRAS.435L..29S,
2017MNRAS.465..242T,
2018ApJ...856..180C}  and all of the known radio magnetars are {intrinsically} almost 100 per cent linearly polarised \citep{2007ApJ...659L..37C,2008ApJ...679..681C,2013Natur.501..391E,2018ApJ...856..180C,2020ApJ...896L..37L,2020MNRAS.498.6044C}. In addition, \citet{agar2021broad} showed that all magnetars have boarder intrinsic pulse profiles (duty cycle more than 4 per cent) compared to most slow pulsars (duty cycle less than 1 per cent).

However, some magnetars contradict these common properties at least for a period of time. For example, PSR~J1846$-$0258, a young radio magnetar, has a rotation period of 0.3 s \citep{2011ApJ...730...66L}, Swift J1818.0-1607 has a steep spectral index of $-$2.8 \citep{2020MNRAS.498.6044C}, and {some magnetars (PSR~J1745$-$2900, XTE~J1810$-$197, 1E~1547.0$-$5408, Swift J1818.0$-$1607) were also reported to have low linear polarisation for some epochs \citep{2007ApJ...659L..37C,2017MNRAS.465..242T,2021MNRAS.502..127L,2023arXiv230207397L}.}

{Finally, nine magnetars have been reported with X-ray quiescent luminosities approximately 20 times lower than the spin-down luminosities, which is unusual for magnetars (1E~1547.0$-$5408, PSR 1622$-$4950, SGR~J1745$-$2900, XTE~J1810$-$197, Swift J1818.0$-$1607, SGR 1833$-$0832, Swift~J1834.9$-$0846, SGR 1935+2154, PSR J1846$-$0258) \citep[see][for reviews]{2014ApJS..212....6O}. Interestingly, six of these are radio loud magnetars, suggesting that radio emission from radio magnetars may have lower X-ray quiescent luminosity.}

PSR~J1746$-$2829 has some observational features commonly found in radio loud magnetars; The spectrum is flatter than most pulsars and two of the radio magnetars. It has a long rotation period. The measured $\dot{P}$ indicates the magnetic field near the lower-end of the magnetar $B_{\rm field}$ limit ($\sim 4.4\times 10^{13}$ G) \citep{2012ApJ...748L..12R}. This study of the radio flux distribution shows that this pulsar has highly variable flux. However, it also shows some properties that are not in common with radio magnetars, but instead are typical for slow pulsars; a narrow pulse profile and low polarisation fraction at high frequency as shown in Figure \ref{fig:lum}. Critically, we have not been able to conclude whether PSR~J1746$-$2829 belongs to the family of magnetars or not.

A magnetar or not, this pulsar is the third flat spectrum pulsar out of seven pulsars found around 0.5$^{\circ}$ from the GC. The other objects are the GC magnetar and a transient flat spectrum pulsar, with a high $B_{\rm{field}}$, named PSR J1746$-$2850, which is another magnetar-like object \citep{2017MNRAS.468.1486D}. If this pulsar, along with PSR J1746$-$2850, is a magnetar, the proportion of magnetars  to non-recycled pulsars in the observed population (0.4 per cent) appears to be significantly different to that of the GC (40 per cent). This could be explained by a larger intrinsic population of magnetars or the environmental conditions in the GC favour the detection of radio magnetars rather than non-recycled pulsars. This could be explained by a higher intrinsic population of magnetars in the GC as predicted by \citet{2014ApJ...783L...7D}, or that the interstellar medium conditions within the GC favour the detection of radio magnetars.

\section{Conclusions}
\label{Conclutions}
We report on the reprocessing of the HTRU-S LowLat around the GC using the first FFA pipeline with an acceleration search. The survey resulted in the discovery of a new slow pulsar with a very high period derivative, 
and a flat spectrum, indicating that this pulsar might be a magnetar. The observations showed that this pulsar has a high DM, RM, and $\tau_{\rm ts}$, which is expected for an object found inside a dense magneto-ionic environment such as the GC. As this pulsar is a magnetar or a high $B_{\rm field}$ pulsar, is the third such object out of seven pulsars located less than 0.5$^{\circ}$ from the GC and may suggest that the GC hosted an anomalously high number of magnetar. All known pulsars detected by the previous FFT processing were detected, as well as a missing pulsar from the FFT based processing of the survey. By comparing the S/N from the FFT to the S/N from the FFA, we found that the FFA pipeline has shown more sensitivity than the FFT pipeline in the searched parameter space.

\section*{Acknowledgements}
\par The Parkes (Murriyang) radio telescope is part of the Australia Telescope, which is funded by the Commonwealth Government for operation as a National Facility managed by CSIRO. This publication is based on observations with the 100-m telescope of the Max-Planck-Institut fuer Radioastronomie at Effelsberg. The localisation of the source was completed using the MPIfR FBSUE backend on the 64-dish MeerKAT radio telescope owned and operated by the South African Radio Astronomy Observatory, which is a facility of the National Research Foundation, an agency of the Department of Science and Innovation. We would like to thank F. Abbate for useful discussions and M. Cruces for providing help for the Effelsberg observations.

\section*{Data Availability}
Pulsar data taken for the P860 and P1050 project are made available through the CSIRO’s Data Access Portal (https://data.csiro.au) after an 18 month proprietary period. The rest of the data underlying this article will be shared on reasonable request to the corresponding author.


\bibliographystyle{mnras}
\bibliography{example} 








\bsp	
\label{lastpage}
\end{document}